\documentclass[12pt,a4paper]{article}
\usepackage{amsmath}
\usepackage{amssymb}
\begin{document}

\begin{titlepage}
\begin{flushright}
IFUP--TH 2007/20\\
\end{flushright}
~

\vskip .8truecm
\begin{center}
\Large\bf
Reduced Hamiltonian for intersecting shells
\footnote{
Work supported in part by M.U.R.S.T.}
\end{center}

\vskip 1.6truecm
\begin{center}
{Francesco Fiamberti} \\
{\small\it Dipartimento di Fisica, Universit{\`a} di Milano and }\\
{\small\it INFN, Sezione di Milano, via Celoria 16, I-20133}\\
{\small\it e-mail: francesco.fiamberti@mi.infn.it}\\
\end{center}
\vskip .8truecm
\begin{center}
{Pietro Menotti} \\
{\small\it Dipartimento di Fisica, Universit{\`a} di Pisa and}\\
{\small\it INFN, Sezione di Pisa, Largo B. Pontecorvo 3, I-56127}\\
{\small\it e-mail: menotti@df.unipi.it}\\
\end{center}
                                                                               
\vskip 1.2truecm
\centerline{August 2007}
                
\vskip 1.2truecm
                                                              
\begin{abstract}

The gauge usually adopted for extracting the reduced Hamiltonian of a thin
spherical shell of matter in general relativity, becomes singular when dealing
with two or more intersecting shells. We introduce here a more general class
of gauges which is apt for dealing with intersecting shells. As an application
we give the hamiltonian treatment of two intersecting shells, both massive and
massless. Such a formulation is applied to the computation of the
semiclassical tunneling probability of two shells. The probability for the
emission of two shells is simply the product of the separate probabilities
thus showing no correlation in the emission probabilities in this model.

\end{abstract}

\vskip 1truecm

\end{titlepage}

\section{Introduction}

A lot of work has been done in the subject of thin spherical shells of
matter in general relativity \cite{israel,hajicek} and several applications
given \cite{polchinski,guth,balbinot}. 
An interesting application of the mechanics of thin shells has been the 
semiclassical treatment of the black hole radiation
\cite{krauswilczek,krauswilczek2,parikhwilczek}.   
An important role in such a treatment is the choice of the gauge. In the
original treatment \cite{krauswilczek} a gauge was adopted in which the
radial component of the metric is equal to the radial coordinate except for an
arbitrarily small region around the shell. Such a gauge choice was put in
mathematically clear terms in \cite{FLW,LWF} where the reduced
canonical momentum is extracted through a well defined limiting process.
However such a limit gauge becomes singular when two or more shells are present
and intersect. In addition also in the simple instance in which only one shell
is present, it would be nice to have a procedure in which no limit process is
present as the gauge choice should be completely arbitrary provided it is free
of coordinate singularities. Moreover even in the one shell case the
procedure for extracting the canonical momentum is usually considered as a very
complicated one. Here we shall give a treatment which
greatly simplifies the derivation of the reduced action, does
not require any limiting procedure and can be applied for the
treatment of two or more shells which intersect. 

All the problem is to derive
the reduced action, i.e. an action in terms of the shell coordinates and some
appropriate conjugate momenta. We shall keep the formalism for the two shell 
case as close as possible to the one shell treatment.

We shall perform all the treatment for a massive
dust shell and the simpler case of a null shell can be derived as a
particular case.  As an application of the developed formalism we
rederive the well known Dray-'t Hooft and Redmount relations for the
intersection of light-like shells \cite{DtH,redmount}. 

The main motivation which
originated the works \cite{krauswilczek,parikhwilczek,parikh} is the
computation of the  
semiclassical tunneling amplitude, which is related to the black hole
radiation in which one takes into account the effect of energy loss by the
black hole. It is remarkable that such a simple model reproduces all the
correct features of the Hawking radiation, giving also some
corrections due to energy conservation. For approaches more kinematical in
nature see e.g.\cite{zerbini}. The adoption of more general gauges allows
the dynamical treatment of two intersecting shells without
encountering singularities. The formalism  developed here is applied to the
computation of the semiclassical emission 
probability of two shells; it was in fact suggested \cite{krauswilczek} that
correlations could show up in the multiple shell emission. We find however
that such a model gives no correlation among the probabilities of the two
emitted shells.

The paper is structured as follows. 

In Section 2 we lay down the formalism by exploiting some peculiar properties
of a function $F$ strictly related to the momenta canonically conjugate to
the metric functions.  It is then just a simple matter of partial integrations
to extract the reduced canonically conjugate momentum without employing any
limit process. This is done in Section 3. A new term containing the time
derivative of the mass of the remnant black hole appears; if such a mass is
considered as a datum of the problem the result agrees with those of the
original massless case of Kraus and Wilczek and with the massive case result
obtained through a limiting process in \cite{FLW}. Then we discuss the
derivation of the equations of motion. In \cite{krauswilczek} and \cite{FLW}
the variational principle was applied by varying the total mass of the system
which we shall denote by $H$. In \cite{parikhwilczek} the attitude was adopted
of keeping the total mass of the system fixed and varying instead the mass of
the remnant black hole.  Here we show that both procedures can be applied to
obtain the equations of motion; depending however on the choice of the gauge
one procedure is far more complicated than the other and we give them both.

In Section 4 we discuss more general gauge choices and derive the equation of
motion in the inner gauge. We shall not consider in the present paper complex
gauges or complex gauge transformations \cite{chowd}.

In Section 5 we discuss the analytic properties of the conjugate momentum;
this is of interest because the imaginary part of the conjugate momentum is
responsible for the tunneling amplitude and in determining the Hawking
temperature both via a simple mechanical model or more precisely by working
out the semiclassical wave functions on which to expand the matter quantum
field \cite{keskivakkuri}. We remark that the result for the tunneling
probability is independent of the mass of the shell but depends only on the
initial and final energy of the black hole.
 
In Section 6 we extend the treatment to two shells in the outer gauge writing
down the explicit expression of the two reduced canonical momenta. 

In Section 7 we derive the equations of motion for both shells from the
reduced two shell action. For simplicity this is done for the massless case;
the general treatment for massive shells is given in Appendix C.

In Section 8 the developed formalism is applied to give a very simple
treatment 
of two shells which intersect. In the massless case we rederive the well 
known
relations of  Dray and 't Hooft \cite{DtH} and Redmount \cite{redmount}.

In Section 9 we consider the problem of computing the tunneling probability
for the emission of two shells which in the process can intersect. To this end
one has to compute the imaginary part of the action along the analytically
continued solution of the equations of motion. In this connection a helpful
integrability result is proven which allows to compute the action along a
specially chosen trajectory on the reduced coordinate space. Such a result
allows on one hand to prove the independence of the result from the
deformation defining the gauge and on the other hand it allows to compute
explicitly such imaginary part. The final outcome is that in all instances
(massive or massless shells) the result again depends only on the initial and
final values of the masses of the black hole and the expression coincides with
the one obtained in the one shell case. The interest in studying the two 
shell system was pointed out in \cite{krauswilczek} in order to
investigate possible correlations among the emitted shells. Here we simply
find that the two shells, even if they interact with an exchange of masses, 
are
emitted with a probability which is simply the product of the probabilities
for single shell emissions and thus that the model does not predict any
correlation among the emitted shells.

In Section 10 we summarize the obtained results.

\section{The action}\label{theactionsec}

As usual we write the metric for a spherically symmetric configuration
in the ADM form \cite{krauswilczek,polchinski,FLW}
\begin{equation}
ds^2=-N^2 dt^2+L^2(dr+N^r dt)^2+R^2d\Omega^2.
\end{equation}

We shall work on a finite region of space time $(t_i, t_f)\times( r_0,
r_m)$ . On the two initial and final surfaces we give the intrinsic
metric by specifying $R(r,t_i)$ and $L(r,t_i)$ and similarly
$R(r,t_f)$ and $L(r,t_f)$.

The complete action in hamiltonian form, boundary terms included is
\cite{hawkinghunter,polchinski,FLW}
\begin{equation}\label{completeaction}
S=S_{shell}+\int_{t_i}^{t_f} dt 
\int_{r_0}^{r_m} dr (\pi_L \dot L +
\pi_R \dot R - N{\cal H}_t-N^r {\cal H}_r) +
\int_{t_i}^{t_f} dt \left.(-N^r \pi_L L+ \frac{NRR'}{L})\right|^{r_m}_{r_0}
\end{equation}
where
\begin{equation}\label{shellaction}
S_{shell}=\int_{t_i}^{t_f} dt ~\hat p~\dot{\hat r}. 
%- N(\hat r,t)
%\sqrt{\frac{\hat p^2}{L^2}+m^2}+N^r(\hat r,t) \hat p\right).
\end{equation}
${\cal H}_t$ and ${\cal H}_r$ are the constraints which are reported in
Appendix A, $\hat r$ is the shell position and $\hat p$ its canonical conjugate
momentum. Action (\ref{completeaction}) is immediately generalized to a finite
number of shells.  
$S_{shell}$ as given by eq.(\ref{shellaction}) refers to a dust shell even
though generalizations to more complicated equations of state have been
considered \cite{hajicek,polchinski,goncalves}.

Varying the action w.r.t. $N$ and $N^r$ gives the vanishing of the constraints 
while
the variations w.r.t. $R,L,\pi_R,\pi_L$ give the equations of motion of
the gravitational field \cite{polchinski, FLW} which for completeness
are reported in Appendix A. The functions $R, L, N, N^r$ are continuous in $r$
while $R', L', N', {N^r}',\pi_L, \pi_R$ can have finite discontinuities 
at the shell position \cite{FLW}. In \cite{FLW} it was proven that the
equation of motion of a massive shell cannot be obtained from a true
variational procedure, in the sense that the obtained expression for
$\dot {\hat p}$ is discontinuous at $r=\hat r$. In the same paper it
was remarked that for consistency the average value of the r.h.s. of the
equation for $\dot{\hat p}$ has to be taken.  For the reader's convenience we
give in Appendix A the
explicit proof that the equations of motion of matter i.e. $\dot{\hat
r}$ and $\dot {\hat p}$ can be deduced from the already obtained equations of
motion for the gravitational field combined with the constraints.
The equation for $\dot{\hat r}$ does not pose any
problem while for the equation for $\dot{\hat p}$ one deduces algebraically
that the r.h.s. contains the 
average of the derivatives of $L,N, N^r$ across the shell. In Appendix A we 
also show directly that such discontinuity is absent in the massless case 
\cite{LWF}.

Already in \cite{polchinski,krauswilczek} it was pointed out that in the region
free of matter, as a consequence of the two constraints the quantity
${\cal M}$
\begin{equation}  
{\cal M}=\frac{\pi_L^2}{2R} +\frac{R}{2}-\frac{R R'^2}{2L^2}
\end{equation}  
is constant in $r$ and this allows to solve for the momenta
\cite{polchinski,krauswilczek}
\begin{equation}
\pi_L=R\sqrt{\left(\frac{R'}{L}\right)^2-1+\frac {2{\cal M}}{R}}\equiv R W
\end{equation}
\begin{equation}
\pi_R= \frac{L[(R/L)(R'/L)'+(R'/L)^2-1+{\cal M}/R]}{W}.
\end{equation}
In the one shell problem we shall call the value of such ${\cal M}$,
$M$ for $r<\hat r$ and $H$ for $r>\hat r$. 
The function 
\begin{equation}
F=R L \sqrt{\left(\frac{R'}{L}\right)^2 -1+ \frac{2{\cal M}}{R}} +
RR'\log\left(\frac{R'}{L}- \sqrt{\left(\frac{R'}{L}\right)^2 -1+ 
\frac{2{\cal M}}{R}}\right)- 
R'~f'(R)
\end{equation}
has the property of generating the conjugate momenta as follows
\begin{equation}
\pi_L = \frac{\partial F}{\partial L}
\end{equation}
\begin{equation}\label{piRfromF}
\pi_R = \frac{\delta F}{\delta R}=\frac{\partial F}{\partial
R}-\frac{\partial }{\partial r}\frac{\partial F}{\partial R'}. 
\end{equation}
The total derivative $\frac{\partial f(R)}{\partial r}= R' f'(R)$ 
of the arbitrary function $f(R)$ does not contribute to the momenta. 
The function $F$ will play a major role in the subsequent developments.
A large freedom is left in the choice of the gauge. With regard to $L$
we shall adopt the usual gauge $L=1$. 
It will be useful in the following developments to choose the
arbitrary function $f(R)$ such that $F=0$ for $R'=1$. Such a
requirement fixes $f'(R)$ uniquely. 
\begin{equation}
F(R,R',{\cal M})
=R W(R,R',{\cal M}) +
RR'({\cal L}(R,R',{\cal M}) - {\cal B}(R,{\cal M}))
\end{equation}
where
\begin{equation}\label{Wdefinition}
W(R,R',{\cal M})= \sqrt{R'^2-1+\frac{2{\cal M}}{R}};~~
{\cal L}(R,R',{\cal M}) = \log(R'-W(R,R',{\cal M}))
\end{equation}
and
\begin{equation}
{\cal B}(R,{\cal M}) = \sqrt{\frac{2{\cal M}}{R}}+\log(1-\sqrt{\frac{2{\cal
M}}{R}}). 
\end{equation}
The function $F(R,R',{\cal M})$ has the following useful properties
\begin{equation}\label{Fproperties}
\frac{\partial F(R,R',{\cal M})}{\partial R'}= R({\cal L}-{\cal B});~~~~
\frac{\partial {\cal L}}{\partial R'}=-\frac{1}{W}.
\end{equation}
Other properties of $F$ will be written when needed.

In the following section we shall choose $R=r$ for $r> \hat r$ and 
also $R=r$ for
$r<\hat r -l$ so that $F$ vanishes identically outside the interval $\hat
r-l<r<\hat r$.  We shall call this class of gauges ``outer
gauges''. In Sect. \ref{moregeneralgauge} we shall consider more general 
gauges e.g. the gauge  $R=r$ for $r< \hat r$ and also $R=r$ for $r>\hat r +l$ 
which we shall call ``inner 
gauges''. However contrary to what is done in \cite{krauswilczek,FLW} we will
not take any limit $l\rightarrow 0$ and prove that the results are
independent of the deformation of $R$ in the region $\hat r-l<r<\hat
r$ (or $\hat r< r<\hat r + l$ for the inner gauges) provided $R'$
satisfies the constraint at $r=\hat r$. We shall call these regions
$(\hat r -l,\hat r)$ for the outer gauge and $(\hat r ,\hat r+l)$ for
the inner gauge, the deformation regions.

The variation of $S$ which produces the equations of motion has to be
taken, as it is well known, by  keeping the metric and in particular
$R$ and $L$ fixed at the boundaries. The variation of the boundary
terms gives
\begin{equation}
-N^r(r_m)\delta \pi_L(r_m)+ N^r(r_0)\delta \pi_L(r_0).
\end{equation}
The $N,~N^r$ can be obtained from the two equations of motion for the
gravitational field
\begin{equation}\label{NrNequations}
0= N[\frac{\pi_L}{R^2}-\frac{\pi_R}{R}] +(N^r)';~~~~
\dot R =- N \frac{\pi_L}{R} +N^r R'.
\end{equation}
Using these it is easily proved that for $r$ outside the deformation
region $(\hat r-l,\hat r)$ we have $N^r=N\sqrt{\frac{2H}{r}}$ for
$r>\hat r$, $N={\rm const}$ and $N^r=N\sqrt{\frac{2M}{r}}$ for
$r<\hat r-l$, $N={\rm const}$ where the two constants as a rule differ. 
Thus the variation of the boundary term is
\begin{equation}
-N(r_m) \delta H+N(r_0) \delta M.
\end{equation}
In the next section we shall connect $N(r_m)$ with $N(r_0)$ being
$N(r)$ not constant in the deformation region.

\section{The one shell effective action in the outer gauge}\label{oneshell} 

As outlined in the previous section we shall choose
\begin{equation}\label{Rfunction}
R(r,t) = r+\frac{V(t)}{\hat r(t)} \int_0^r \rho(r'-\hat r(t)) dr'=
r+\frac{V(t)}{\hat r} g(r-\hat r(t))
\end{equation}
having $\rho$ support $(-l,0)$, $\rho(0)=1$ and smooth in $-l$ and 
\begin{equation}
\int_{-l}^{0} \rho(r) dr =0.
\end{equation}
As a consequence the deformation $g(r)$ has support in $(-l,0)$ 
and $g'(0-\varepsilon)=1$. Such $R$ satisfies the
discontinuity requirements at $r=\hat r$ which are imposed by the 
constraints.
In fact the constraints (see Appendix A) impose the following discontinuity
relations at $\hat r$ (we recall that we chose $L\equiv 1$)
\begin{equation}\label{DeltaR1}
\Delta R'=-\frac{V}{R};~~~~V=\sqrt{{\hat p}^2+m^2}
\end{equation}
and
\begin{equation}\label{DeltapiL}
\Delta \pi_L=-\hat p.
\end{equation}

In the outer gauge the bulk gravitational action becomes
\begin{equation}\label{bulkaction}
S_g = \int_{t_i}^{t_f} I_g ~dt
\end{equation}
where, keeping in mind that $L\equiv 1$
\begin{eqnarray}
&& I_g=\int_{r_0}^\infty (\pi_L \dot L +\pi_R \dot R ) dr=\int_{r_0}^\infty
\pi_R \dot R dr= 
\int_{r_0}^\infty
\left((\frac{\partial F}{\partial
R}-\frac{\partial }{\partial r}\frac{\partial F}{\partial R'})\dot R
\right) dr \nonumber \\
&& =\int_{r_0}^{\hat r(t)} \frac{dF}{dt}dr-\dot M(t)\int_{r_0}^{\hat
r(t)}\frac{\partial F}{\partial M} dr-
\left.\frac{\partial F}{\partial R'}
\dot R\right |_{r_0}^{\hat r(t)} \nonumber \\
&& =\frac{d}{dt }\int_{r_0}^{\hat r(t)} F dr- \dot M(t)\int_{r_0}^{\hat
r(t)}\frac{\partial F}{\partial M} dr -\left[\dot{\hat r}(t) F
+\frac{\partial F}{\partial R'}
\dot R\right]_{\hat r(t)-\varepsilon}
\end{eqnarray}
where we used the fact that $F$ vanishes at $r=r_0$.

Adding $I_{shell} = \hat p \dot{\hat r}$ and neglecting the total time
derivative  which does not contribute to the equations of motion, we obtain
for the reduced action in the outer gauge 
\begin{equation}\label{outerreducedaction}
\int_{t_i}^{t_f} \left(p_{c}~ \dot{\hat r} -\dot
M(t)\int_{r_0}^{\hat r(t)}\frac{\partial F}{\partial M} dr+ 
\left.(-N^r \pi_L + NRR')\right|^{r_m}_{r_0}\right)dt
\end{equation}
where using (\ref{DeltaR1},\ref{DeltapiL})
\begin{eqnarray}\label{pc}
&& p_c= - F(\hat r(t)-\varepsilon) - \frac{1}{\dot {\hat
r}(t)}\left.\frac{\partial F}{\partial R'}\dot R \right|_{\hat
r(t)-\varepsilon} +\hat p = \nonumber \\
&& = \sqrt{2M\,\hat r}-\sqrt{2H \,\hat r}-\hat
r\log\left(\frac{\hat r+\sqrt{{\hat p}^2+m^2}-\hat p-
\sqrt{2H \,\hat r}}{\hat r-\sqrt{2M \,\hat r}}\right).
\end{eqnarray}
A few comments are in order: 1) No limit $l\rightarrow 0$ is necessary
for obtaining $p_c$ of eq.(\ref{pc}) which holds for any deformation
$g$. 2) The $\dot M(t)$ term is important, as we shall see, if we
consider the variational problem in which $M$ is varied. On the other
hand if we consider $M$ as a datum of the problem and vary $H$ the
contribution $\dot M(t)$ is absent. 3) $\hat p$ in eq.(\ref{pc}) is a
function of $\hat r$, $H$ and $M$ as given by the discontinuity
relation (\ref{DeltapiL}) equivalent to the implicit equation
\begin{equation}\label{fundamentalH}
H-M= V +\frac{m^2}{2\hat r}-\hat p\sqrt{\frac{2H}{\hat r}}.
\end{equation}
We discuss now these issues in more detail. Let us consider at first $M$
as a datum of the problem and vary $H$. As shown in Appendix A in order
to be consistent with the gravitational equations $M$ has to be
constant in time. This is the situation examined in \cite{FLW} where
the expression (\ref{pc}) for $p_c$ was derived by a limit process in
which $l\rightarrow 0$. From eq.(\ref{outerreducedaction}) 
we see that the equation of
motion for $\dot{\hat r}$ is given by
\begin{equation}\label{outereqmotion}
\dot{\hat r}\frac{\partial p_c}{\partial H}-N(r_m)=0
\end{equation}
where $N(r_m)$ can be replaced by $N(\hat r)$ being $N(r)$ in the
outer gauge constant for $r>\hat r$. The computation of 
eq.(\ref{outereqmotion})
keeping in mind the implicit definition of $\hat p$
eq.(\ref{fundamentalH}) 
gives
the correct equation of motion for the massive shell
\begin{equation}\label{1sheqmotion}
\dot{\hat r} = \frac{\hat p}{V}N(\hat r)- N^r(\hat r) =
\left(\frac{\hat p}{V}-\sqrt{\frac{2H}{\hat r}}\right)N(\hat r).
\end{equation}
The outline of the calculation is done in Appendix B.

Alternatively one can consider $H={\rm const}$ as a datum of the
problem and vary $M(t)$. We remark that as shown in Appendix A the
datum $M$ or $H$ is consistent with the gravitational equations only if
$H$ and $M$ are constant in time. Nevertheless in the variational
problem $H$ and $M$ have to be considered as functions of time, because
the constraints tell us only that $M$ and $H$ are constant in
$r$. Only after deriving the equation of motion one can insert the
consequences of the gravitational equations of motion.

The variation of $M(t)$ in the outer gauge is a far more complicated
procedure, due to 
the presence of $\dot M(t)$, but 
gives rise to the same result obtained by varying $H$ and keeping $M$
fixed. As this will be useful to understand the two shell reduced
dynamics to be developed in Sect.\ref{twoshell} 
we go into it with some detail.  
In this case the $\dot M$ term plays
a major role; in fact the equation of motion now takes the form
\begin{equation}\label{Meq}
\dot{\hat r} \frac{\partial p_c}{\partial
M}+\frac{d}{dt}\int_{r_0}^{\hat r(t)}\frac{\partial F}{\partial M} dr 
+ N^r(r_0)\frac{\partial \pi_L}{\partial M}(r_0) =0
\end{equation}
where due to the vanishing of $g(x)$ for $x<-l$ we have
$\pi_L(r_0)=\sqrt{2Mr_0}$. $N^r(r)$ on the other hand is obtained by
solving the two coupled equations (\ref{NrNequations}) with the 
condition that for $r>\hat r$, $N^r(r)$ equals $\sqrt{\frac{2H}{r}}$, having
normalized $N=1$ for $r>\hat r$. 
One easily finds that for $r<\hat r$ 
\begin{equation}\label{Nrsolution}
N^r(r)= W \left[\int_{\hat r}^r \dot R \frac{\partial
\pi_R}{\partial M}dr + \frac{\sqrt{2H\hat r}}{\sqrt{2H\hat r}+\hat p}\right]. 
\end{equation}
Taking into account that
\begin{equation}
-\frac{d}{d r}(\frac{\partial^2 F}{\partial R'\partial M})
+\frac{\partial^2 F}{\partial M\partial R}=
\frac{\partial\pi_R}{\partial M}
\end{equation}
we have
\begin{equation}
\frac{d}{dt}\int_{r_0}^{\hat r(t)}\frac{\partial F}{\partial M} dr =
\int_{r_0}^{\hat r}\dot R\frac{\partial \pi_R}{\partial M}dr 
+\left[\dot{\hat
r}\frac{\partial F}{\partial
M}+\frac{\partial^2F}{\partial M\partial R'}\dot R\right]_{\hat r-\varepsilon} 
\end{equation}
and eq.(\ref{Meq}) becomes
\begin{equation}\label{generalMeqmotion}
\dot{\hat r}\frac{\partial p_c}{\partial M}+\left(\dot{\hat r} \frac{\partial
F}{\partial M}
+ \dot R\frac{\partial^2 F}{\partial R'\partial M}\right)_{\hat
r-\varepsilon}+\frac{\sqrt{2H\hat r}}{\sqrt{2H\hat r}+\hat p}=0. 
\end{equation}
From the expression for $p_c$ given by the first line of
eq.(\ref{pc}) 
we obtain
\begin{equation}
\dot{\hat r}\left(\frac{\partial \hat p}{\partial
M}-\left.\frac{V}{W}\frac{\partial R'}{\partial M}\right|_{\hat r
-\varepsilon}\right)+\frac{\sqrt{2H\hat r}}{\sqrt{2H\hat r}+\hat p}=0 
\end{equation}
where $W(\hat r -\varepsilon)=\sqrt{R'^2(\hat r-\varepsilon)-1+2M/R(\hat r)} 
=\sqrt{\frac{2H}{\hat r}}+\frac{\hat p}{\hat r}$
and $R'(\hat r-\varepsilon)=1+V/\hat r$. 
Using
\begin{equation}
\frac{\partial \hat p}{\partial M}=-\frac{1}{\frac{\hat
p}{V}-\sqrt{\frac{2H}{\hat r}}}
\end{equation}
we obtain eq.(\ref{1sheqmotion}) again. We remark once more that no
limit process $l\rightarrow 0$ is necessary for all these developments.

To summarize, in the present section we have derived the reduced action for
the one shell problem in the outer gauge with an arbitrary
deformation. One can vary $H$ (the exterior ADM mass) considering the
interior mass as given, or one can vary the interior mass $M$
considering the exterior mass $H$ as given, or even one can vary both
$M$ and $H$ always obtaining the correct equations of motion. Whenever
$M$ is varied the $\dot M$ term in eq.(\ref{Meq}) plays a 
crucial role. All the results do not depend on the deformation $g$.

\bigskip

\section{\bf More general gauges}\label{moregeneralgauge} 

It is of interest to examine more general gauges given by eq.(\ref{Rfunction})
where 
$g$ does not necessarily vanish for positive argument, i.e. we can consider
$g(x)$ with $g(x)=0$ for $|x|>l$, $g(0)=0$ and $g'(+0)-g'(-0)=-1$, 
thus again satisfying
the constraint (\ref{DeltaR1}). 
In this case $F(R,R',H)$ does not vanish identically
for $r>\hat r $ and the bulk action (\ref{bulkaction}) is given by the time
integral of 
\begin{equation}\label{generalgaugebulk}
I_g=\frac{d}{dt }\int_{r_0}^{r_m} F dr- \dot M(t)\int_{r_0}
^{\hat r(t)}\frac{\partial F}{\partial M} dr- \dot H(t)\int_{\hat r(t)}^{r_m}
\frac{\partial F}{\partial H} dr +\left[\dot{\hat r}(t) F
+\frac{\partial F}{\partial R'}
\dot R\right]_{\hat r(t)-\varepsilon}^{\hat r(t)+\varepsilon}
\end{equation}
The $p_c$ is easily computed using  eq.(\ref{generalgaugebulk}) and one gets
immediately 
the general form for the canonical momentum $p_c$
\begin{equation}
p_c = \hat r (\Delta {\cal L}-\Delta {\cal B})
\end{equation}
where $\Delta {\cal L}={\cal L}(\hat r+\varepsilon)-{\cal L}(\hat
r-\varepsilon)$ and similarly for $\Delta {\cal B}$ and the reduced action
becomes
\begin{equation}\label{genreducedaction}
S = \int_{t_i}^{t_f} \left(p_{c}~ \dot{\hat r} -\dot
M(t)\int_{r_0}^{\hat r(t)}\frac{\partial F}{\partial M} dr -\dot
H(t)\int_{\hat r(t)}^{r_m}\frac{\partial F}{\partial H} dr+ 
\left.(-N^r \pi_L + NRR')\right|^{r_m}_{r_0}\right ) dt.
\end{equation}
We shall call inner gauge the one characterized by $g(x)=0$ for $x<0$. 

Due to the similarity with the treatment of Sect.\ref{oneshell} 
we shall go
through rather quickly. Now $F$ vanishes identically for $r<\hat r$ 
and the action takes the form
\begin{equation}
S=\int_{t_i}^{t_f}dt\left (p_c^i \dot{\hat r}-\dot H\int_{\hat
r(t)}^{r_m}
\frac{\partial F}{\partial H}dr+(-N^r\pi_L+N R R')|^{r_m}_{r_0}\right) 
\end{equation}
where $p_c^i$ is given by
\begin{equation}
p_c^i = \left[F+ \frac{\dot R}{\dot{\hat r}}\frac{\partial F}{\partial
R'}\right]_{\hat r +\varepsilon}+\hat p
\end{equation}
whose explicit value is
\begin{equation}
p_c^i= \sqrt{2M\,\hat r}-\sqrt{2H \,\hat r}-\hat
r\log\left(\frac{\hat r-\sqrt{2H \,\hat r}}
{\hat r- V + \hat p-\sqrt{2M\hat r}}\right)
\end{equation}
and $\hat p$, again determined by the discontinuity equation
(\ref{DeltapiL}), is given by the implicit equation
\begin{equation}\label{fundamentalM}
H-M=V-\frac{m^2}{2\hat r}-\hat p \sqrt{\frac{2M}{\hat r}}
\end{equation}
which is different from eq.(\ref{fundamentalH}).
%showing the gauge dependence of
%$\hat p$ and $p_c$. 
Now the simple procedure is the one in which one varies
$M$  keeping $H$ as a fixed datum of the problem.  
This time the solution of the system eq.(\ref{NrNequations}) 
gives simply $N={\rm const}$ and $N^r = N \sqrt{\frac{2M}{r}}$ for 
$r_0<r<\hat r$.

\bigskip

\section{\bf The analytic properties of $p_c$}\label{analyticmomenta}

We saw that in the outer gauge 
\begin{equation}\label{pcanalytic}
p_c= \sqrt{2M\,\hat r}-\sqrt{2H \,\hat r}-\hat
r\log\left(\frac{\hat r+V-\hat p-
\sqrt{2H \,\hat r}}{\hat r-\sqrt{2M \,\hat r}}\right).
\end{equation}
The solution of eq.(\ref{fundamentalH}) for $\hat p$ is
\begin{equation}
\frac{\hat p}{\hat r} =\frac{A \sqrt{\frac{2H}{\hat r}}~
\pm\sqrt{A^2 -(1-\frac{2H}{\hat r})\frac{m^2}{\hat r^2}}}
{1-\frac{2H}{\hat r}}
\end{equation}
where
\begin{equation}
A =\frac{H-M}{\hat r}-\frac{m^2}{2 \hat r^2}.
\end{equation}
If we want $\hat p$ to describe an outgoing shell we must choose the
plus sign in front of the square root. Moreover the shell
reaches $r=+\infty$ only if $H-M>m$ as expected. 

The logarithm has branch points at zero and infinity and thus we must
investigate for which values or $\hat r$ such values are reached.
At $\hat r=2H$, $\hat p$ has a simple pole with positive residue; then the
numerator goes to zero and below $2H$ it becomes
\begin{equation}\label{contnumerator}
1 - \frac{V}{\hat r}-\frac{\hat p}{\hat r}-\sqrt{\frac{2H}{\hat r}}
\end{equation}
where here $V$ is the absolute value of the square root. Expression 
(\ref{contnumerator}) is negative irrespective of the sign of $\hat p$ and
stays so 
because $\hat p$ is no longer singular.
In order to compute the tunneling amplitude, below $\hat r = 2H$ we
have to use the prescription \cite{parikhwilczek} $\hat r - 2H
\rightarrow \hat r - 2H - i\varepsilon$ and as a consequence the $p_c$
below $2H$ acquires the imaginary part $i\pi \hat r$. Below $\hat r =
2M$ the denominator of the argument of the logarithm in 
eq.(\ref{pcanalytic})
becomes negative so that the argument of the logarithm reverts to positive
values. 
Thus the ``classically forbidden'' region is $2M<\hat r <2H$
independent of $m$ and of the deformation $g$ and the integral of the 
imaginary part of $p_c$ for any deformation $g$ is
\begin{equation}\label{integratedimpart}
\int {\rm Im}~ p_c dr = 
\pi \int_{2M}^{2H} r dr = 2\pi(H^2-M^2)= 4\pi (M+\frac{\omega}{2})\omega
\end{equation}
with $\omega = H-M$ which is the Parikh-Wilczek result \cite{parikhwilczek}.
A more profound way to relate the result for $p_c$ to the formula for the
Hawking radiation is to use (\ref{pc}) to compute semiclassically the modes on
which to expand the quantum field, and then proceed as usual by means of the
Bogoliubov transformation. This was done in \cite{krauswilczek,keskivakkuri}.
A more direct particle like interpretation of (\ref{integratedimpart}) was
given e.g. in \cite{hamilton}.

Similarly one can discuss the analytic properties of the
conjugate momentum $p_c^i$ in the inner gauge. We have
\begin{equation}\label{pcright}
p_c^i =\sqrt{2M\hat r} -\sqrt{2H\hat r} - 
\hat r \log\left(\frac{1-\sqrt{\frac{2H}{ \hat r}}}
{1 -\frac{V}{\hat r} -\sqrt{\frac{2M}{\hat r}}+\frac{\hat p}{\hat
r}} \right).
\end{equation}
This time the solution of eq.(\ref{fundamentalM}) gives for $\hat p$
\begin{equation}\label{hatpi}
\frac{\hat p}{\hat r} =
\frac{\sqrt{\frac{2M}{\hat r}}~A\pm\sqrt{A^2 -(1-\frac{2M}{\hat
r})\frac{m^2}{\hat r^2}}}{1 -\frac{2 M}{\hat r}}
\end{equation}
where now
\begin{equation}
A=\frac{H-M}{\hat r} +\frac{m^2}{2\hat r^2}.
\end{equation}
Notice that for $m=0$ we have $p_c=p^i_c$. To describe an outgoing shell the 
square root in (\ref{hatpi}) has to be taken with the positive sign.

All the point is the discussion of the sign of the term
\begin{equation}
1 -\frac{V}{\hat r} -\sqrt{\frac{2M}{\hat r}}+\frac{\hat p}{\hat r}=
R'(\hat r + \varepsilon)-\sqrt{R'^2(\hat r + \varepsilon)-1+\frac{2H}{\hat r}}
\end{equation}
where
\begin{equation}\label{W}
\sqrt{\frac{2M}{\hat r}}-\frac{\hat p}{\hat r}=
\sqrt{R'^2(\hat r + \varepsilon)-1+\frac{2H}{\hat r}}.
\end{equation}
For $m=0$ at $\hat r=2H$ eq.(\ref{W}) is negative so that at $\hat r = 2H$
\begin{equation}\label{denominator}
1 -\frac{V}{\hat r} -\sqrt{\frac{2M}{\hat r}}+\frac{\hat p}{\hat r}=
R'(\hat r + \varepsilon)+\sqrt{R'^2(\hat r + \varepsilon)-1+\frac{2H}{\hat r}}
\end{equation}
is positive, being in eq.(\ref{denominator}) the square root on the
r.h.s. understood as the positive determination of the square root. 
The same happens for $m\neq 0$ provided $m<H-M$ which is the
condition for the shell to be able to reach $\hat r =+\infty$. For $\hat r<2H$
as  $-1+\frac{2H}{\hat r}>0$ such a term stays positive irrespective of the 
sign of $R'$, until $R'$ diverges. This happens at $2M$ where $\hat p$ given
by 
eq.(\ref{hatpi}) has a 
simple pole with positive residue. Thus
below $2M$ eq.(\ref{denominator}) reverts to
\begin{equation}
1 -\frac{V}{\hat r} -\sqrt{\frac{2M}{\hat r}}+\frac{\hat p}{\hat r}=
R'(\hat r + \varepsilon)-\sqrt{R'^2(\hat r + \varepsilon)-1
+\frac{2H}{\hat r}}
\end{equation}
which is negative irrespective of the sign of $R'(\hat r + \varepsilon)$. 
The conclusion is that
$p^i_c$ takes the imaginary part $i\pi \hat r$ in the interval $2M, 2H$ as it
happens for $p_c$.

\bigskip

\section{\bf The two shell reduced action}\label{twoshell}

From now on we shall work in the outer gauge.
We denote with ${\hat r}_1$ and ${\hat r}_2$ the coordinates of
the first and second shell $\hat r_1 <\hat r_2$. The value of ${\cal
M}$ for $r <\hat r_1$ will be denoted by $M$, for $\hat r_1 <r<\hat
r_2$ by $M_0$ and for $r>\hat r_2$ will be denoted by $H$ as before. In
extending the treatment to two interacting shells we shall keep the
formalism as close as possible to the one developed in
Sect. \ref{oneshell}. 
The most relevant difference is that in any gauge the intermediate mass $M_0$
and the total mass $H$ always intervene dynamically. We shall consider the
mass $M$ as a datum of the problem.

For the metric component $R$ we shall use
\begin{equation}\label{twoshellR}
R(r) = r+v_2 g(r-\hat r_2)+v_1 h(r-\hat r_1);~~~~
v_2= \frac{V_2}{R(\hat r_2)}; ~~~~v_1 =\frac{V_1}{R(\hat r_1)}
\end{equation}
where $h(x)$ has the same properties of $g(x)$ described in
Sect. \ref{oneshell} (actually we could use the same function). 
Both $g$ and
$h$ vanish for positive argument and thus we are working in an outer
gauge according to the definition of Sect. \ref{oneshell}.

The action is given by the time integral of
\begin{equation}
I =\hat p_2 \dot{\hat r}_2+\hat p_1 \dot{\hat r}_1 +\int dr \pi_R \dot R
+ b.t.
\end{equation}
Breaking the integration range from $r_0$ to $\hat r_1$ and from
$\hat r_1$ to $\hat r_2$ and using eq.(\ref{piRfromF}) for $\pi_R$ and the
same technique as used for the one shell case, we reach the expression
\begin{eqnarray}\label{twoshellF}
&& \hat p_2 \dot{\hat r}_2+\hat p_1 \dot{\hat r}_1+
\frac{d}{dt}\int_{r_0}^{\hat r_2} F dr
-\dot M_0\int^{\hat r_2}_{\hat r_1}\frac{\partial F}{\partial M_0}dr
-\dot{\hat r}_2 F(\hat r_2-\varepsilon) - \left(\dot R\frac{\partial
F}{\partial R'}\right)(\hat r_2-\varepsilon)+\nonumber\\
&& \dot{\hat r}_1\Delta F(\hat r_1) + \Delta\left(\dot R \frac{\partial F}
{\partial R'}\right)(\hat r_1)\\
&& =\hat p_1 \dot{\hat r}_1 +\frac{d}{dt}\int_{r_0}^{\hat r_2} F
dr-\dot M_0\int^{\hat r_2}_{\hat r_1}\frac{\partial F}{\partial M_0}dr
+p_{c2}^0~ \dot{\hat r}_2+ \dot{\hat r}_1~\Delta F(\hat r_1) +
\Delta\left(\dot R \frac{\partial F}{\partial R'}\right)(\hat r_1) \nonumber
\end{eqnarray}
where $p_{c2}^0$ is given by eq.(\ref{pc}) with $M$ replaced by $M_0$, and 
$\Delta$ stays for the jump
across the discontinuity at $\hat r_1$. 
We recall that $F$ depends only on $R$,$R'$ and ${\cal M}$ and the
partial derivative w.r.t. these variables will have the usual
meaning. For the other quantities appearing in the calculations we
recall that the independent variables are $M$, $M_0$, $H$, $\hat r_1$,
$\hat r_2$ and when taking partial derivatives we shall consider the 
remaining variables as fixed. 
In eq.(\ref{twoshellF}) we have, using eq.(\ref{Fproperties}), denoting with
$\Delta$ the discontinuity across $\hat r_1$ and with the bar the average at
$\hat r_1$
\begin{eqnarray}
&& \dot{\hat r}_1\hat p_1+\dot{\hat r}_1\Delta F
+\Delta\left(\frac{\partial F}{\partial 
R'}\dot R\right)= \\
&& =\dot{\hat r}_1\hat p_1+\dot{\hat r}_1\Delta F+ 
\overline{\frac{\partial F}{\partial
R'}}~\Delta\dot R + \overline{\dot R}\Delta\frac{\partial
F}{\partial R'}= \nonumber \\
&&\dot{\hat r}_1[\overline{R'}R( \Delta {\cal L} -
\Delta{\cal B})+\Delta R' R( \bar{\cal L} -\bar{\cal B})]+\Delta \dot R
~R(\bar{\cal L}-\bar{\cal B})
+\overline{\dot R}R( \Delta{\cal L} -\Delta{\cal B})=\nonumber
\end{eqnarray}
\begin{equation}
= \dot{\hat r}_1p_{c1} + \dot{\hat
r}_2\tilde p_{c2}+ \dot H (R(\hat r_1)-\hat r_1)\frac{\partial T}{\partial
H}{\cal D} + \dot M_0 (R(\hat r_1)-\hat r_1)\frac{\partial T}{\partial M_0}
{\cal D}\nonumber
\end{equation}
where
\begin{equation}\label{pc1}
T=\log v_2;~~~~
{\cal D} = R (\Delta{\cal L}- \Delta{\cal B});~~~~
p_{c1}= R'(\hat r_1+\varepsilon){\cal D};
\end{equation}
\begin{equation}\label{tildepc2}
\tilde p_{c2}= -(R'(\hat r_1+\varepsilon)-1){\cal D}+(R(\hat r_1)-\hat
r_1)\frac{\partial T}{\partial \hat r_2}{\cal D}=
\frac{d}{d\hat r_2}(R(\hat r_1)-\hat r_1){\cal D}
\end{equation}
having used
\begin{equation}
\Delta \dot R = - \dot{\hat r}_1 \Delta{R'}
\end{equation}
and
\begin{equation}
\bar{\dot R} = -\dot{\hat r}_1\frac{v_1}{2}-
\dot{\hat r}_2(\bar{R'}-1)+
(R(\hat r_1)-\hat r_1)\frac{dT}{dt}.
\end{equation}
Summing up the reduced action for the two shell system is given,
boundary terms included, by the time integral of
\begin{eqnarray}\label{twoshellreducedaction}
&& \dot{\hat r}_1p_{c1} + \dot{\hat
r}_2p_{c2}+ \dot H (R(\hat r_1)-\hat r_1)\frac{\partial T}{\partial
H}{\cal D} + \dot M_0 (R(\hat r_1)-\hat r_1)\frac{\partial T}{\partial M_0}
{\cal D}+ \nonumber\\
&& +\frac{d}{dt}\int_{r_0}^{\hat r_2} F dr-\dot M_0\int^{\hat r_2}_{\hat
r_1}\frac{\partial F}{\partial M_0}dr 
+(-N^r\pi_L+N R R')|^{r_m}_{r_0} 
\end{eqnarray}
where $p_{c2}= p_{c2}^0+\tilde p_{c2}$ with $p^0_{c2}$ given by
eq.(\ref{pc}) with $M$ replaced by $M_0$ and $\tilde p_{c2}$ by
eq. (\ref{tildepc2}).  
With regard to eq.(\ref{twoshellreducedaction}) we notice that  irrespective
of the gauge used both 
terms in $\dot M_0$ and $\dot H$ appear in the action. Moreover $p_{c1}$ and
$p_{c2}$ depend both on $\hat r_1$ and $\hat r_2$.

\bigskip

\section{The two shell equations of motion}

From the reduced action (\ref{twoshellreducedaction}) we can derive
the equations of motion for the two shells. This is of some importance
in order to show the consistency of the scheme. We recall that action
(\ref{twoshellreducedaction}) has been derived in the outer gauge
i.e. $R(r)=r$ for $r>\hat r_2$  ($\hat r_2>\hat r_1$). While in the
one shell problem formulated in the outer gauge $\dot H$ does not
appear, in the two shell problem it is always present. Again we consider
$M={\rm const.}$ as a datum of the problem.

In the variational procedure we can vary $\hat r_1$, $\hat r_2$, $H$
and $M_0$ independently. We shall start varying $H$ but keeping all
other parameters fixed. For the sake of simplicity we shall deal
here with the massless case $m_1=m_2=0$. In Appendix C we give the
general derivation for $m_1$ and $m_2$ different from zero. The most important
fact that occurs when we 
vary $H$ keeping $M_0$ fixed is that the terms proportional to
$\dot{\hat r}_1$ cancel in the variation. The simplifying feature of
the massless case is that $\Delta{\cal L}=0$ so that in eq.(\ref{pc1})
${\cal D}= -R\Delta {\cal B}$, being ${\cal B}$ function only of $R$ and
${\cal M}$ and not of $R'$ and thus
only of $R$ and $M_0$ being $M$ a datum of the problem. The
coefficient of $\dot{\hat r}_1$, taking into account the following relation, 
easily
derived from (\ref{Fproperties}), 
\begin{equation}\label{dFsdM}
\frac{\partial F}{\partial {\cal M}}
+R' R \frac{\partial {\cal B}}{\partial {\cal M}}=\frac{1}{W- R'}
\end{equation}
is found proportional to
\begin{equation}\label{dotr1coefficient}
\frac{\partial}{\partial H}(R'(\hat r_1+\varepsilon) {\cal D})-
(R'(\hat r_1+\varepsilon)-1) \frac{\partial T}{\partial H} {\cal D}-
(R(\hat r_1)-\hat r_1) \frac{\partial T}{\partial H} 
\frac{\partial {\cal D}}{\partial R} R'(\hat r_1+\varepsilon)
\end{equation}
where we used
\begin{equation}
\frac{d R(\hat r_1)}{dt} = \dot{\hat r}_1 R'(\hat r_1+\varepsilon)
-\dot{\hat r}_2 (R'(\hat r_1+\varepsilon)-1)
+\frac{dT}{dt}(R(\hat r_1)-\hat r_1)
\end{equation}
and we took into account that $T$ does not depend on $\hat r_1$ and that on
the equations of motion $\dot H=\dot M_0=0$.
Now employing the relations
\begin{equation}\label{derivativerelations}
\frac{\partial R(\hat r_1)}{\partial H}=\frac{\partial T}{\partial H}
(R(\hat r_1)-\hat r_1);
~~~~\frac{\partial R'(\hat r_1+\varepsilon)}{\partial H}=
\frac{\partial T}{\partial H} (R'(\hat r_1+\varepsilon)-1)
\end{equation}
we see that expression (\ref{dotr1coefficient}) vanishes. Thus we are
left only with the $\dot{\hat r}_2$ terms. We know already that the
boundary term and the $p^0_{c2}$ term give the correct equation of
motion and thus we have simply to prove that the $\dot{\hat r}_2$
terms originating from
\begin{equation}
-\frac{d}{dt}[(R(\hat r_1)-\hat r_1)\frac{\partial T}{\partial H}{\cal D}]
\end{equation}
cancel $\displaystyle{\frac{\partial \tilde{p}_{c2}}{\partial H}}$ i.e.
\begin{equation}\label{dotr2add}
\frac{\partial}{\partial H}[-(R'(\hat r_1+\varepsilon)-1){\cal D}+(R(\hat
r_1)-\hat r_1)\frac{\partial T}{\partial \hat r_2}{\cal D}]-
\frac{\partial}{\partial \hat r_2}[(R(\hat
r_1)-\hat r_1) \frac{\partial T}{\partial H}{\cal D} ]=0. 
\end{equation}
Using relations (\ref{derivativerelations}) we have that eq.(\ref{dotr2add})
is satisfied. Such a result is expected as the exterior shell
parameterized by $\hat r_2$ moves irrespective of the dynamics which
develops at lower values of $r$ until $\hat r_1$ crosses $\hat r_2$.

Now we vary $M_0$ keeping $H$ fixed. We have no boundary
term contribution because also $M$ is constant and we find the
equation
\begin{eqnarray}
&& \dot{\hat r}_1\frac{\partial p_{c1}}{\partial M_0}+
\dot{\hat r}_2\frac{\partial p_{c2}}{\partial M_0}
-\frac{d}{dt}[(R(\hat r_1)-\hat r_1)\frac{\partial T}{\partial M_0}{\cal D}]
+\dot{\hat r}_2\frac{\partial F(\hat r_2-\varepsilon)}{\partial M_0}-
\dot{\hat r}_1\frac{\partial F(\hat r_1+\varepsilon)}{\partial M_0}+
\nonumber\\
&& +\int^{\hat r_2}_{\hat r_1}\dot R\frac{\partial\pi_R}{\partial M_0}dr
+\left.\dot R\frac{\partial^2F}{\partial M_0\partial R'}\right|^{\hat
r_2-\varepsilon}_{\hat r_1+\varepsilon}=0. 
\end{eqnarray}
Using expression (\ref{Nrsolution}) for $N^r(r)$,  
relation (\ref{NrNequations}) for $N(r)$ and relation (\ref{dFsdM})
we obtain with $\alpha = \sqrt{2H\hat r}/(\sqrt{2H\hat r}+\hat p)$
\begin{eqnarray}\label{firstdothatr1}
&& [R'(\hat r_1+\varepsilon)- W(\hat r_1+\varepsilon)]^{-1}
(\dot{\hat r}_1 + N^r(\hat r_1)-N(\hat r_1)) \nonumber\\
&& +\dot{\hat r}_2 \frac{\partial p^0_{c2}}{\partial M_0}+
\left[\dot{\hat r}_2 \frac{\partial F}{\partial M_0}+
\frac{\partial^2F}{\partial M_0\partial R'}\dot R\right]_{\hat
r_2-\varepsilon} +\alpha
\nonumber\\
&&-[R'(\hat
r_1+\varepsilon)- W(\hat r_1+\varepsilon)]^{-1}
\frac{\dot R}{W}(\hat
r_1+\varepsilon) 
\nonumber\\
&& +\dot{\hat r}_2\frac{\partial \tilde p_{c2}}{\partial M_0}
-\left.\frac{\partial^2F}{\partial M_0\partial R'}\dot R\right|_{\hat 
r_1+\varepsilon} -\dot{\hat r}_2\frac{\partial }{\partial \hat r_2} [(R(\hat
r_1)-\hat r_1)\frac{\partial T}{\partial M_0}{\cal D}] =0.
\end{eqnarray}
The first line is the equation of motion for $\hat r_1$, the second line
vanishes being the equation of motion for $\hat r_2$ (see
eq.(\ref{generalMeqmotion}) of the
one shell problem) while exploiting the relation
\begin{equation}
\frac{\partial^2 F}{\partial {\cal M}\partial R'}+R\frac{{\partial \cal B}}
{\partial {\cal M}}=\frac{1}{W(W-R')}
\end{equation}
also derived from (\ref{Fproperties})
and substituting into $\dot{\hat r}_2$ the value given by the equations of
motion derived previously we find that the third line simplifies
with the fourth. Thus we are left with the relation 
\begin{equation}
\dot{\hat r}_1 + N^r(\hat r_1)-N(\hat r_1)=0
\end{equation}
which is the correct equation of motion for the interior shell.

\section{Exchange relations}

In the present section we consider in the above developed formalism the
intersection of two shells during their motion. In the massless case
we shall rederive the well known relations of Dray and 't Hooft \cite{DtH} 
and Redmount \cite{redmount} between the
masses characteristic of 
the three regions before and after the collision. We shall denote by
$t_0$ the instant of collision, $\hat r_0$ the coordinate of the
collision. It is well known that some hypothesis has to be done on the
dynamics of the collision which should tell us which are the masses of the 
two shells after the collision. Once these are given the problem is reduced to
that of a two particle collision in special relativity. The formulas which we 
develop 
below give the intermediate mass $M'_0$ after the collision. The simplest
assumption is that of the ``transparent crossing'' i.e. after the collision
shell $1$ carries along with unchanged 
mass $m_1$ and the same for shell $2$. A further simplification occurs in
the massless case i.e. when the two massless shells go over to two massless
shells. We develop first the general formalism which applies also when we
have a change in the masses during the collision and then specialize to
particular situations. 

Just before the collision i.e. at $t=t_0-\varepsilon$  assuming that
$\hat r_1(t_0-\varepsilon)<\hat r_2(t_0-\varepsilon)$ we have for the
momentum $\pi_L$
\begin{eqnarray}
&& \pi_L(\hat r_1-0,t_0-\varepsilon)=\pi_L(\hat
r_1+0,t_0-\varepsilon)+\hat p_1= \\
&& =\pi_L(\hat
r_2-0,t_0-\varepsilon)+\hat p_1 = \pi_L(\hat
r_2+0,t_0-\varepsilon)+\hat p_1+\hat p_2 \nonumber  
\end{eqnarray}
and after the collision i.e. at  $t=t_0+\varepsilon$ we have with $\hat
r_2(t_0+\varepsilon) <\hat r_1(t_0+\varepsilon)$, 
\begin{eqnarray}
&& \pi_L(\hat r_2-0,t_0+\varepsilon)=\pi_L(\hat
r_2+0,t_0+\varepsilon)+\hat p'_2= \\
&& \pi_L(\hat
r_1-0,t_0+\varepsilon)+\hat p'_2 = \pi_L(\hat
r_1+0,t_0+\varepsilon)+\hat p'_1+\hat p'_2 \nonumber  
\end{eqnarray}
and at the collision $\hat r_2=\hat r_1=\hat r_0$. The main point in
treating the crossing is to realize that the sum $V_1+V_2$ has to be
continuous in time as it represents the discontinuity of $R'$ at the
time of crossing. In fact just before the crossing we have for $r<\hat
r_0$, choosing for simplicity in (\ref{twoshellR}) $g(x)=h(x)$, 
\begin{equation}
R(r) = r+ \frac{(V_1+V_2)}{\hat r_0}g(r-\hat r_0)
\end{equation}
while immediately after we have
\begin{equation}
R(r) = r+ \frac{(V'_1+V'_2)}{\hat r_0}g(r-\hat r_0).
\end{equation}
If $V_1+V_2\neq V'_1+V'_2$ we would have a discontinuous evolution
of the metric for $r<\hat r_0$ which is incompatible with the equations
of motion of the gravitational field. As $M$ is unchanged during the
time evolution $V_1+V_2= V'_1+V'_2$ implies that $\pi_L(\hat
r_1-0,t_0-\varepsilon)=\pi_L(\hat
r_2-0,t_0+\varepsilon)$ which combined with the fact that
$\pi_L(\hat r_2+0,t_0-\varepsilon)=\pi_L(\hat
r_1+0,t_0+\varepsilon)=\sqrt{2H \hat r_0}$ gives the further
relation $\hat p_1+\hat p_2=\hat p'_1+\hat p'_2$. Thus we have the same
kinematics as in the two particle collision in special relativity. In the case 
of conservation of the masses $m_1$ and $m_2$ we have the same kinematics
as that  of an elastic collision in $1+1$ dimensions. A discussion of special
cases with massive shells using a different formalism has been given in
\cite{nunez}. 

The relation between $\pi_L(\hat r_1-0, t_0-\varepsilon)$ 
and $\pi_L(\hat r_2+0,t_0-\varepsilon)$
\begin{equation}
\hat r_0\sqrt{\left(1+\frac{V_1+V_2}{\hat
r_0}\right)^2-1+\frac{2M}{\hat r_0}}=\sqrt{2H\hat r_0}+\hat p_1+\hat p_2
\end{equation}
gives the equation
\begin{equation}\label{generalexchange}
m_1^2+m_2^2+2(V_1V_2-\hat p_1\hat p_2)+2(V_1+V_2)\hat r_0+2M\hat r_0=
2H \hat r_0+2(\hat p_1+\hat p_2)\sqrt{2H\hat r_0}
\end{equation}
where $M_0$ is given by
\begin{equation}\label{fundamentalM0}
H-M_0 = V_2+\frac{m_2^2}{2\hat r_0}-\hat p_2\sqrt{\frac{2H}{\hat r_0}}.
\end{equation}
From the knowledge of ${\hat p}'_1$ one derives $M_0'$ from 
\begin{equation}\label{fundamentalM10}
H-M'_0 = V'_1+\frac{{m'_1}^2}
{2\hat r_0}-{ \hat p'}_1\sqrt{\frac{2H}{\hat r_0}}.
\end{equation}
For ``transparent'' crossing i.e. $m'_j=m_j$ and ${\hat p}'_j= {\hat p}_j$, 
$M'_0$ is given by (\ref{fundamentalM10})
with ${\hat p}'_1={\hat p}_1$.
The massless case is most easily treated. In this case in order for
the shells to intersect they must move in opposite directions, the
exterior one with negative and the interior one with positive velocity. 
The conservation of $V_1+V_2$ and $\hat p_1+\hat p_2$ in the massless
case has two solutions which are physically equivalent i.e. $\hat p'_1
= \hat p_1$, $\hat p'_2 = \hat p_2$. The two shells intersect only if
$\hat p_2 <0$ and $\hat p_1>0$ so that $V_2=-\hat p_2$ and $V_1=\hat
p_1$. 
From eq.(\ref{fundamentalM0}) we have
\begin{equation}\label{hatp20}
\hat p_2 =-\frac{H-M_0}{1+\sqrt{\frac{2H}{\hat r_0}}}
\end{equation}
and from $\hat p'_1 = \hat p_1$
\begin{equation}\label{hatp10}
\hat p_1 =\frac{H-M'_0}{1-\sqrt{\frac{2H}{\hat r_0}}}.
\end{equation}
Eq.(\ref{generalexchange}) now becomes
\begin{equation}
H \hat r_0+\sqrt{2H\hat r_0}(\hat p_1+\hat p_2)= \hat
r_0(\hat p_1-\hat p_2)+ M \hat r_0-2\hat p_1\hat p_2
\end{equation}
and substituting here eqs.(\ref{hatp20},\ref{hatp10})
we obtain
\begin{equation}
H \hat r_0 + M \hat r_0-2 H M = M_0 \hat r_0- 2 M_0 M'_0 + M'_0 \hat r_0
\end{equation}
which is the well known Dray- 't Hooft and Redmount relation
\cite{DtH,redmount}.

\section{Integrability of the form $p_{c1}d\hat r_1+p_{c2}d\hat r_2$ and the
two shell tunneling probability}

We are interested in computing the action for the two
shell system i.e. the integral 
\begin{equation}\label{intpc1pc2}
\int_{t_i}^{t_f} dt (p_{c1}~\dot{\hat r}_1+ p_{c2}~\dot{\hat r}_{c2})=
\int_{r_{1i}, r_{2i}}^{r_{1f}, r_{2f}} (p_{c1}~d\hat r_1+p_{c2}~d\hat r_2)
\end{equation}
on the solution of the equations of motion.
This is of interest in the computation of the semiclassical wave
function and the tunneling probability in the two shell case.
We shall prove explicitly that the form $p_{c1}d\hat r_1+p_{c2}d\hat
r_2$ is closed i.e. integrable. Such a result 
will be very useful in the actual computation and in showing the
independence of the results of the deformation $g$. 
In fact we can reduce the computation to the integral on a simple path 
on which the two momenta $p_{c1}$ and $p_{c2}$ take a simpler form.

The integrability result is similar to a theorem of analytical
mechanics \cite{whittaker,arnold} stating that for a system with two degrees of
freedom, in 
presence of a constant of motion the form $p_1 dq_1+p_2 dq_2$ where
the $p_j$ are expressed in terms of the energy and of the constant of
motion, is a closed form. Here the setting is somewhat different as we
are dealing with an effective action with two degrees of freedom,
arising from the action of a constrained system. We recall that for
$\hat r_1<\hat r_2$ $M_0$ like $H$ is a constant of motion in virtue of
the equations of motion of the gravitational field. The $p_{cj}$ are
functions of $H,M_0,\hat r_1,\hat r_2$ in addition to the fixed datum
of the problem $M$. The effective action takes the form
\begin{equation}
\int dt( p_{c1}\dot{\hat r}_1+p_{c2}\dot{\hat r}_2 + p_H \dot H+
p_{M_0}\dot M_0) + b.t.
\end{equation}
where $b.t.$ is the boundary term (see eq.(\ref{completeaction})) 
which depends only on
$H, N^r$ and not 
on $\hat r_j$. The value of $N^r$ is supplied by the solution of the 
gravitational 
equations of motion. Thus varying w.r.t. $\hat r_1$ we have
\begin{equation}
0 = -\frac{dp_{c1}}{dt}+\dot{\hat r}_1 \frac{\partial p_{c1}}{\partial
\hat r_1} +\dot{\hat r}_2\frac{\partial p_{c2}}{\partial\hat r_1} 
+\frac{\partial p_H}{\partial
\hat r_1}\dot H +\frac{\partial p_{M_0}}{\partial
\hat r_1} \dot M_0.
\end{equation}
We show in Appendix A that the constraints combined with the
equations 
for the gravitational field impose $0=\dot H=\dot M_0$
and thus we have
\begin{equation}\label{integrabilitycond}
\frac{\partial p_{c1}}{\partial \hat r_2}-\frac{\partial
p_{c2}}{\partial \hat r_1} =0.
\end{equation}
The meaning of the procedure is that the consistency of the
variational principle imposes eq.(\ref{integrabilitycond}). On the other hand
eq.(\ref{integrabilitycond}) can 
be verified also from the explicit expression of $p_{c1}$ and $p_{c2}$ given
in Sect. \ref{twoshell}.
If the crossing of the shells occurs outside the region $2M, 2H$ we
can choose the path as to keep the two deformations non overlapping in such a
region; then $p_{c1}$ takes the simple form (\ref{pc}) with $H$ substituted by
$M_0$ and $p_{c2}$ again the form (\ref{pc}) with $M$ substituted by $M_0$ 
and thus one obtains for the imaginary part of
integral (\ref{intpc1pc2})
\begin{equation}\label{oneptimpart}
\frac{1}{2}\left((2 M_0)^2-(2 M)^2+(2 H)^2-(2 M_0)^2\right)=
\frac{1}{2}\left((2 H)^2-(2 M)^2\right)
\end{equation}
i.e. the two shells are emitted independently.
If the crossing occurs at $\hat r_0$ with $2M<\hat r_0<2H$, with e.g. $\hat
r_1<\hat r_2$ before the crossing, we choose the path $\hat r_1 = \hat
r_2-\varepsilon$ before the crossing and  $\hat r_2 = \hat
r_1-\varepsilon$ after the crossing.
For clearness sake we examine at first the problem for the crossing of a null 
shell with a massive shell, even when the mass of the massive shell changes.
In order to reduce the integration path to the described one, one must first,
given the two initial points with $R(\hat r_{1i})< 2M $ and $R(\hat r_{2i})<2M
$, bring them together ( $\hat r_1$ will denote the position of the massless
shell).  
For $\hat r_{1i}< \hat r_{2i}$ this is done by integrating
along the line $\hat r_2 = {\rm const}$ with $\hat r_1$ varying from $\hat
r_{1i}$ to $\hat r_{2i}-\varepsilon$. The contribution is
\begin{equation}\label{coalesce} 
\int_{\hat r_{1i}}^{\hat r_{2i}-\varepsilon} p_{c1} d\hat r_1=
\int_{\hat r_{1i}}^{\hat r_{2i}-\varepsilon} R'(\hat r_1+\varepsilon) ~ {\cal
D}~ d\hat r_1. 
\end{equation}
But $R'(\hat r_1)$ is real and
\begin{equation}\label{calDm0} 
{\cal D} = - R(\hat r_1) \left[\sqrt{\frac{2M_0}{R(\hat r_1)}}+
\log\left(\frac{1-\sqrt{\frac{2M_0}{R(\hat r_1)}}}
{1-\sqrt{\frac{2M}{R(\hat r_1)}}}\right)
-\sqrt{\frac{2M}{R(\hat r_1)}}\right].
\end{equation}  
For $R(\hat r_1)<2M<2M_0$ eq.(\ref{calDm0}) is real and thus the contribution
(\ref{coalesce}) is real. 
If on the other hand $\hat r_{2i}<\hat r_{1i}$ we integrate $p_{c1}$ from
$\hat r_{1i}$ to $\hat r_{2i}+\varepsilon$ keeping $\hat r_2$ fixed. This time
we recall that
\begin{equation}
p_{c1}= p^0_{c1}+\tilde p_{c1} 
\end{equation}
where $p^0_{c1}$ is real because we are outside the interval $(2M_0, 2H)$, and
\begin{equation}\label{pc1m0}
\tilde p_{c1}= -(R'(\hat r_2+\varepsilon)-1) {\cal D}(\hat r_2)+
(R(\hat r_2)-\hat r_2)\frac{\partial T}{\partial \hat r_1} {\cal
D}(\hat r_2) 
\end{equation}
with $T$ now given by $\log v_1$. Again all the items in eq.(\ref{pc1m0}) are
real. 
Thus we can start e.g. with $\hat r_{1i} = \hat r_{2i}-\varepsilon$.
The integration along the path $\hat r_1 = \hat r_2 -\varepsilon$ up to $\hat
r_0$ is very simple because 
\begin{equation}
p_{c2}+p_{c1}=\hat r\left[\sqrt{\frac{2M}{\hat r}}-\sqrt{\frac{2H}{\hat r}}-
\log\frac{1+\frac{V_2-\hat p_2}{\hat r}-\sqrt{\frac{2H}{\hat
r}}}{1-\sqrt{\frac{2M}{\hat r}}}\right]
\end{equation} 
whose complete discussion has already been done in
Sect. \ref{analyticmomenta}.  
The ``gap'' is
$2M,2H$. More difficult is the analysis for $\hat r_0<\hat r <2H$. Now
we have
\begin{equation}\label{p2primeplsup1prime}
p'_{c2}+p'_{c1}=\hat r\left[\sqrt{\frac{2M}{\hat r}}-\sqrt{\frac{2H}{\hat r}}-
\log\frac{1+\frac{V'_2-\hat p'_2}{\hat r}-\sqrt{\frac{2H}{\hat
r}}}{1-\sqrt{\frac{2M}{\hat r}}}\right]
\end{equation}
with
\begin{equation}
\hat {p}'_1=\frac{H-M'_0}{1-\sqrt{\frac{2H}{\hat r_1}}}.
\end{equation}
Moreover
\begin{equation}\label{hatp2prime}
\frac{\hat{p}'_2}{\hat r_2}
=\frac{AW(\hat r_2+\varepsilon)+R'(\hat r_2+\varepsilon)
\sqrt{A^2-(1-\frac{2M'_0}{\hat
r_2})\frac{m_2^2}{\hat{r}_2^2}}}{1-\frac{2M'_0 }{\hat r_2}}
\end{equation}
where
\begin{equation}
A = \frac{M'_0-M}{\hat r_2}-\frac{m_2^2}{2{\hat r}^2_2}
\end{equation}
and
\begin{equation}
W(\hat r_2+\varepsilon) = \sqrt{(1+\frac{\hat p'_1}{\hat
r_2})^2-1+\frac{2M_0}{\hat r_2}}.
\end{equation}
${\hat p}'_1$ diverges with positive residue at $\hat r_1 = 2 H$ and thus 
also $W(\hat r_2+\varepsilon)$ and ${\hat p}'_2$ of eq.(\ref{hatp2prime}) 
diverge like $\hat p'_1$. As a consequence the analytic continuation of
$V'_2-{\hat p}'_2$ below $\hat r = 2 H$ is negative which makes the numerator
of the argument of the logarithm in eq.(\ref{p2primeplsup1prime}) negative. 
We show now that such numerator stays negative all the way for $\hat r<2H$.
The argument of the square root in (\ref{hatp2prime}) never vanishes
so that at $\hat r_2 = 2M'_0$ the numerator in (\ref{hatp2prime}) reduces to
\begin{equation}
A(W(\hat r_2+\varepsilon)+R'(\hat r_2+\varepsilon)).
\end{equation}
We can now explicitly compute $W(\hat r_2+\varepsilon)$ and 
$R'(\hat r_2+\varepsilon)$ at  $\hat r_2 =
2M'_0$. At such a point we have
\begin{equation}
R'(\hat r_2+\varepsilon) = 1+\frac{(H-M'_0)/\hat r_2}
{1-\sqrt{\frac{2H}{\hat r_2}}}=
\frac{1}{2}\left(1-\sqrt\frac{H}{M'_0}\right)
\end{equation}
while
\begin{equation}
W(\hat r_2+\varepsilon) = \sqrt\frac{2H}{2M'_0}+ \frac{1}{\hat r_2}
\frac{H-M'_0}{1-\sqrt{\frac{2H}{\hat r_2}}}=
\sqrt\frac{H}{M'_0}+ \frac{1}{2}
\left(\frac{\frac{H}{M'_0}
-1}{1-\sqrt{\frac{H}{M'_0}}}\right)=-\frac{1}{2}
\left(1-\sqrt\frac{H}{M'_0}\right)  
\end{equation}
which means that there is no pole in $\hat{p}'_2$ at $\hat r
=2M'_0$. Thus $\hat{p}'_2$ is regular below $2H$ and $V_2'-\hat{p}'_2$
cannot change sign.

In this way we proved that for the crossing of a null shell and a
massive shell, into a null shell and another massive shell even with
a change of mass, the imaginary part of the integral (\ref{intpc1pc2}) is still
given by eq.(\ref{oneptimpart}).

The reasoning in the general case of both $m_1$ and $m_2$ different from zero
is dealt with similarly. It is sufficient to examine the case $\hat r_1<\hat
r_2$ the other case being now equivalent. We shall first discuss the
integration along the path $\hat r_1 +\varepsilon = \hat r_2$ 
up to $\hat r_0$.
For $\hat r_1+\varepsilon=\hat r_2=\hat r$ we have
\begin{equation}\label{pc2pluspc1}
p_{c2}+p_{c1}=\hat r\left[\sqrt{\frac{2M}{\hat r}}-\sqrt{\frac{2H}{\hat r}}-
\log\frac{1+\frac{V_2-\hat p_2 + V_1-\hat p_1}{\hat r}-\sqrt{\frac{2H}{\hat
r}}}{1-\sqrt{\frac{2M}{\hat r}}}\right]
\end{equation}
with
\begin{equation}
H-M_0 = V_2+\frac{m_2^2}{2\hat r_2}-\hat p_2 \sqrt{\frac{2H}{\hat r_2}}
\end{equation}
which as before makes $\hat p_2$ diverge at 
$\hat r_2 = 2 H$. For $\hat p_1$ we have
\begin{equation}\label{genhatp1}
\frac{\hat p_1}{\hat r_1} = \frac{A_1 W(\hat r_1+0) + R'(\hat
r_1+0) \sqrt{A_1^2 - (1-\frac{2M_0}{\hat r_1})\frac{m_1^2}
{\hat r_1^2}}}{1-\frac{2M_0}{\hat r_1}}
\end{equation}
with
\begin{equation}
A_1= \frac{M_0-M}{\hat r_1}-\frac{m_1^2}{2{\hat r}^2_1}.
\end{equation}
$\hat p_2$ diverges with positive residue at $\hat r=2H$ and $\hat p_1$ also
diverges like $\hat p_2$ as
\begin{equation}\label{refeq}
R'(\hat r_1+0)=1+\frac{V_2}{\hat r_1}
\end{equation}
and $W(\hat r_1+0)$ as given by eq.(\ref{Wdefinition}) also diverges like $\hat
p_2$. Then 
for $\hat r$ just below $2H$ we have both $V_2-\hat p_2<0$ and $V_1-\hat
p_1<0$. As before all the point is in proving that $\hat p_1$ is regular
below $2H$ i.e. does not diverge at $\hat r=2M_0$. To this end we must examine
the numerator of eq.(\ref{genhatp1}) at $\hat r = 2M_0$. 
We have for $\hat r=2M_0$
\begin{equation}
W(2M_0+0) = \frac{1}{2}(\sqrt{\frac{H}{M_0}}-1)
+\frac{m_2^2}{8 M^2_0(\sqrt{\frac{H}{M_0}}+1)} >0
\end{equation}
as $H>M_0$.
With regard to $R'(\hat r_1+0) $ which is negative for $\hat r_1=
2H-\varepsilon$ it cannot 
change sign for $2 M_0<\hat r<2 H$ because at the point where $R'(\hat r_1+0)$
vanishes 
\begin{equation}
W(\hat r_1+0) = \sqrt{{R'}^2(\hat r_1+0)-1+\frac{2M_0}{\hat r_1}}
\end{equation}
would become imaginary while from $W(\hat r_1+0)=\sqrt{2H/\hat
r_1} +\hat p_2/\hat r_1$ we have that $W$ is real. Then at $\hat r = 2M_0$ the
numerator in eq.(\ref{genhatp1}) vanishes and $\hat p_1$ 
is regular all the way below $2H$. Thus we are in the same situation as
in the previously discussed case. Below $\hat r = 2 M$ the argument of the
logarithm in eq.(\ref{pc2pluspc1}) becomes positive again due to the change in
sign of the denominator $1-\sqrt{2 M/\hat r}$. The integration for $\hat r >
\hat r_0$ is treated simply by exchanging $m_1$ with $m_2$.

Finally we deal with the integral
\begin{equation}
\int_{\hat r_{2i}}^{\hat r_{1i}+\varepsilon} p_{c2}~ d\hat r_2
\end{equation}
which makes the two initial points coalesce. As in the previously discussed
case of one massless shell, all the point is in proving that $\cal D$ is
real. In addition to the contribution of ${\cal B}$ which we already proved to
be real, we have now the contribution of ${\cal L}$. We have already proven
that at $r=\hat r_2-\varepsilon$
\begin{equation}\label{pc2ext}
R'(r) -\sqrt{R'(r)^2-1+\frac{2M_0}{R(r)}} =R'(r)-W(r)  
\end{equation}
equals
\begin{equation}
1+\frac{V_2}{\hat r_2}-\frac{\hat p_2}{\hat
r_2}-\sqrt{\frac{2H}{\hat r_2}} 	 	
\end{equation}
which is negative for $\hat r_2 < 2H$ and thus in particular for 
$\hat r_2 < 2M$.
In the interval $\hat r_1 <r <\hat r_2<2M $ the term (\ref{pc2ext}) cannot
change sign because $-1+ 2M_0/R$ is positive. Moreover we have
\begin{equation}\label{Lminus}
R'(\hat r_1-\varepsilon)-W(\hat r_1-\varepsilon)  
= R'(\hat r_1+\varepsilon)-W(\hat r_1+\varepsilon)+ \frac{V_1}{R(\hat r_1)}
-\frac{\hat p_1}{R(\hat r_1)}~.   
\end{equation}
But we proved after eq.(\ref{refeq}) that below $2 H$, the analytic
continuation of $V_1-\hat p_1$ is negative, implying that (\ref{Lminus}) is
negative, like $R'(\hat r_1+\varepsilon)-W(\hat r_1+\varepsilon)$. The outcome
is that the discontinuity of ${\cal L}$ at $\hat r_1$, being given by the
logarithm of a positive number, is real. This concludes
the proof in the case of the emission of two massive shells.

\section{Conclusions}

The main issue of the present paper is the treatment of two intersecting
shells of matter or radiation in general relativity,  
in a formalism,
apt to compute the tunneling amplitude for the emission of two shells.
In order to do so it is necessary to adopt a gauge which is more general than
the one used \cite{krauswilczek,FLW} in the treatment of a single shell. In
the usual treatments of 
the tunneling amplitude for a single shell, a limit gauge is
adopted. Already at the level of a single shell we show that the complete
action contains a term in which the mass of the remnant black hole plays a
dynamical role. Such a term is unimportant if the variation of the action is
taken with respect to the total mass of the system, keeping the mass of the
remnant as a datum of the problem, but becomes essential if one varies the mass
of the remnant keeping the total mass of the system fixed as done e.g. in
\cite{parikhwilczek}. The reduced canonical momentum even in the single shell
instance is gauge dependent but the tunneling probability turns out to be
independent of such a choice. All the treatment is performed in the general
massive case, the massless one being a special case. The tunneling results 
are
independent of the mass of the shell. The adoption of a general non-limit
gauge allows the extension of the formalism to two or more shells. In
this instance both the intermediate mass and the total mass become dynamical
variables in the sense that the reduced action contains terms proportional to
the time derivative of them. We show how to derive the equations of motion of
both the interior and exterior shell by varying the reduced action and this is
done both in the massless and in the more complicated massive case. With
regard to 
the computation of the tunneling amplitude it is possible to prove an
integrability theorem which allows to deform the trajectory in coordinate
space to a contour which drastically simplifies the computation. Firstly one
proves in such a way that the result is independent of the deformation
defining the gauge 
introduced in Sect. \ref{theactionsec}, and secondly one
finds in the 
general massive or massless case that the tunneling probability is given
simply by the product of the tunneling probabilities for the independent
emission of the two shells. Such a circumstance is interpreted \cite{parikh} as
the fact that in this model we have no information encoded in the radiation
emitted by the black hole. 

\bigskip\bigskip

\section*{Acknowledgments}

One of us (P.M.) is grateful to M. Porrati and V. S. Rychkov for stimulating 
discussions. 

\bigskip

\section*{Appendix A}

For completeness we report here some formulas which are
useful in the text.
The constraints are given by \cite{polchinski,krauswilczek,FLW}
\begin{equation}
{\cal H}_r = \pi_R R'- \pi_L'L -\hat p~\delta(r-\hat r),
\end{equation}
\begin{equation}
{\cal H}_t = \frac{R R''}{L}+\frac{{R'}^2}{2
L}+\frac{L \pi_L^2}{2R^2}-\frac{R R' L'}{L^2}-
\frac{\pi_L\pi_R}{R}-\frac{L}{2}+ \sqrt{{\hat p}^2
L^{-2}+m^2}~\delta(r-\hat r).
\end{equation}
From the constraints it follows \cite{polchinski} that the quantity
\begin{equation}\label{Minvariant}
{\cal M} = \frac{\pi_L^2}{2 R}+\frac{R}{2} -\frac{R\,(R')^2}{2L^2}
\end{equation}
is constant in the regions of $r$ where there are no sources as there
\begin{equation}\label{Mprime}
{\cal M}' = -\frac{R'}{L}{\cal H}_t -\frac{\pi_L}{RL}{\cal H}_r.
\end{equation}
The equations of motion for the gravitational field are \cite{FLW}
\begin{equation}\label{Ldot}
\dot L = N\left(\frac{L\pi_L}{R^2}-\frac{\pi_R}{R}\right)+
(N^r L)',
\end{equation}
\begin{equation}\label{Rdot}
\dot R = -\frac{N\pi_L}{R}+ N^r R',
\end{equation}
\begin{equation}\label{piLdot}
\dot\pi_L
=\frac{N}{2}\left[-\frac{\pi_L^2}{R^2}-\left(\frac{R'}{L}\right)^2+1
+\frac{2~\hat p^2~\delta(r-\hat r)}{L^3\sqrt{\hat p^2 \hat L^{-2}+m^2}}\right]
-\frac{N'R' R}{L^2}+N^r\pi_L',
\end{equation}
\begin{equation}\label{piRdot}
\dot\pi_R
=N\left[\frac{L\pi_L^2}{R^3}-\frac{\pi_L\pi_R}{R^2}-\left(\frac{R'}{L}\right)'
 \right]-\left(\frac{N'R}{L}\right)'+(N^r\pi_R)'.
\end{equation}
The equations of motion for $\dot{\hat r}$ and $\dot{\hat p}$ follow
from the above equations for $R$, $L$, $\pi_R$ and $\pi_L$ and the
constraints.
In fact using the relation 
\begin{equation}\label{contR}
\frac{dR(\hat r)}{dt}=\dot R(\hat r+\varepsilon)
+R'(\hat r+\varepsilon) \dot{\hat r}=\dot R(\hat r-\varepsilon)
+R'(\hat r-\varepsilon) \dot{\hat r}
\end{equation}
we have
\begin{equation}
\dot{\hat r}\Delta R'+\Delta \dot R=0.
\end{equation}
Using now the equation of motion (\ref{Rdot}) and the constraints
\begin{equation}
\Delta R'(\hat r) = -\frac{V}{R};~~~~ \Delta \pi_L(\hat r)
=-\frac{\hat p}{L}
\end{equation}
where $V=\sqrt{\hat p^2 + m^2 L^2}$,
one obtains
\begin{equation}\label{hatrdot}
\dot{\hat r} =\frac{N(\hat r)\hat p}{L(\hat r)\sqrt{\hat p^2 
+m^2 L^2(\hat r)}}- N^r(\hat r).
\end{equation}
Using the relation
\begin{equation}\label{contL}
\frac{dL(\hat r)}{dt} = L'(\hat r+\varepsilon) \dot{\hat r}+\dot L(\hat
r+\varepsilon) =L'(\hat r-\varepsilon) \dot{\hat r}+\dot L(\hat
r-\varepsilon)= \overline{L'}\dot{\hat r} +\overline{\dot L} 
\end{equation}
and
\begin{equation}\label{discpiRdot}
-\dot{\hat r} \Delta\pi_R = -N(\hat r)\frac{\Delta R'}{L(\hat r)}
-\frac{\Delta N'~R(\hat r)}{L(\hat r)}+ N^r(\hat r) \Delta\pi_R
\end{equation}
derived from (\ref{piRdot}) and
\begin{equation}
\dot{\hat p} = -\frac{dL(\hat r)}{dt}\Delta\pi_L-L(\hat
r)\frac{d\Delta\pi_L}{dt} 
\end{equation}
one obtains
\begin{equation}\label{hatpdot}
\dot{\hat p} =\frac{N(\hat r) {\hat p}^2 \overline{{L}'}}
{L^2(\hat r)\sqrt{\hat p^2 + m^2 L^2(\hat r)}}- 
\frac{\overline{{N}'}}{L(\hat r)} \sqrt{\hat p^2
+m^2 L^2(\hat r)}+ \hat p ~\overline{({N}^r)'}.
\end{equation}
The occurrence of the average values $\overline{L'}=
[L'(\hat r+\varepsilon)+ L'(\hat r -\varepsilon)]/2$ etc. in the
previous equation, was pointed out
and discussed at the level of the variational principle in
\cite{FLW}. In the massless case $m=0$ however, using again
relations (\ref{contR},\ref{contL},\ref{discpiRdot}) one proves that
the r.h.s. of eq.(\ref{hatpdot}) has no discontinuity i.e. there is
no need to take the average value in the r.h.s. of eq.(\ref{hatpdot}). 
In fact let us consider the discontinuity 
of the zero mass version of the r.h.s. of eq.(\ref{hatpdot})
\begin{equation}
\hat p(\frac{\eta N L'}{L^2}- \frac{\eta N'}{L}+{N^r}')
\end{equation}
being $\eta$ the sign of $\hat p$, i.e.
\begin{equation}\label{discontinuity}
\hat p(\frac{\eta N \Delta L'}{L^2}- \frac{\eta \Delta N'}{L}+
\Delta {N^r}').
\end{equation}
From eq.(\ref{Ldot}) and the equation of motion (\ref{hatrdot}) we have
\begin{equation}
\eta\frac{N}{L}\Delta L' =
\frac{N}{R}\Delta\pi_R-\frac{NL}{R^2}\Delta\pi_L-L\Delta {N^r}'
\end{equation} 
and from eq.(\ref{piRdot})
\begin{equation}
\eta N\Delta\pi_R=R\Delta N'+N\Delta R'.
\end{equation}
Substituting the two into eq.(\ref{discontinuity}) and taking into account that
$\Delta\pi_L=-\hat p/L$ and $\Delta R'=-\eta \hat p/R$ we have that expression
(\ref{discontinuity}) vanishes.
This fact was discussed at the level of the variational principle in 
\cite{LWF}.  
 
From the equations of motion (\ref{Ldot},\ref{Rdot},\ref{piLdot},
\ref{piRdot}) it follows that in the region where there are no sources
\begin{equation}
\frac{d{\cal M}}{dt} = -N\frac{R'}{L^3}{\cal H}_r- N^r
\frac{R'}{L}{\cal H}_t - N^r \frac{\pi_L}{RL} {\cal H}_r
\end{equation}
i.e. combining with (\ref{Mprime}), we have that  in the region free
of sources ${\cal M}$ is constant both in $r$ and in $t$.

\bigskip

\section*{Appendix B}

Equation of motion for $\hat r$. 

1) Outer gauge

In this case 
\begin{equation}
p_c = R(\Delta {\cal L}-\Delta {\cal B}) = \hat r\left(-{\cal L}(\hat
r-\varepsilon) -\sqrt{\frac{2H}{\hat r}}+\sqrt{\frac{2M}{\hat r}}+
\log\left(1-\sqrt{\frac{2M}{\hat r}}\right)\right).
\end{equation}
Using
\begin{equation}
\frac{\partial \hat p}{\partial H}=\left(1+\frac{\hat p}
{\sqrt{2H \hat r}}\right)\left(\frac{\hat p}{V}- \sqrt{\frac{2H}
{\hat r}}\right)^{-1}
\end{equation}
and
\begin{equation}
\frac{\partial {\cal L}}{\partial R'}(\hat r-\varepsilon) = -\frac{1}{W(\hat
r-\varepsilon)} = -\left(\sqrt{\frac{2H}{\hat r}}+\frac{\hat p}
{\hat r}\right)^{-1}
\end{equation}
we have
\begin{equation}
\frac{\partial p_c}{\partial H} = -\sqrt{\frac{\hat r}{2H}}+\sqrt{\frac{\hat
r}{2H}} \frac{\hat p}{V}\frac{1}{\frac{\hat p}{V} -\sqrt{\frac{2H}{\hat r}}}=
\left(\frac{\hat p}{V}-\sqrt{\frac{2H}{\hat r}}\right)^{-1}
\end{equation}
from which
\begin{equation}
\dot{\hat r} = N(r_m) \left(\frac{\hat p}{V}-\sqrt{\frac{2H}{\hat r}}\right).
\end{equation}

2) Inner gauge

This time we have
\begin{equation}
H-M = V -\frac{m^2}{2\hat r} - \hat p \sqrt{\frac{2M}{\hat r}}
\end{equation}
and
\begin{equation}
p^i_c = \hat r\left({\cal L}(\hat
r+\varepsilon) -\sqrt{\frac{2H}{\hat r}}+\sqrt{\frac{2M}{\hat r}}
-\log\left(1-\sqrt{\frac{2H}{\hat r}}\right)\right).
\end{equation}
Using
\begin{equation}
\frac{\partial \hat p}{\partial M}=- \left(1-\frac{\hat p}
{\sqrt{2M \hat r}}\right)\left(\frac{\hat p}{V}- 
\sqrt{\frac{2M}{\hat r}}\right)^{-1}
\end{equation}
and
\begin{equation}
\frac{\partial {\cal L}}{\partial R'}(\hat r+\varepsilon) = -\frac{1}{W(\hat
r+\varepsilon)} = -\left(\sqrt{\frac{2M}{\hat r}}-\frac{\hat p}
{\hat r}\right)^{-1}
\end{equation}
we have
\begin{equation}
\frac{\partial p^i_c}{\partial M} =\sqrt{\frac{\hat r}{2 M}}
\left[1- \frac{\hat p}{V} 
\left(\frac{\hat p}{V}-\sqrt{\frac{2M}{\hat r}}\right)^{-1}\right] =
-\left(\frac{\hat p}{V}-\sqrt{\frac{2M}{\hat r}}\right)^{-1} 
\end{equation}
i.e.
\begin{equation}
\dot {\hat r} = \left(\frac{\hat p}{V} -
\sqrt{\frac{2M}{\hat r}}\right)N(r_0).
\end{equation}

\section*{Appendix C}

In this Appendix we derive the equations of motion for the exterior and
interior shell from the reduced action (\ref{twoshellreducedaction}) for
masses $m_1$ and $m_2$ different from zero.

First we consider the variation $\delta H\neq 0$ and $\delta M_0=0$. Under
such a variation the coefficient of the $\dot{\hat r}_1$ term is given, using
\begin{equation}
\frac{\partial R}{\partial H}=(R-\hat r_1)\frac{dT}{dH};~~~~
\frac{\partial R'}{\partial H}=(R'-1)\frac{dT}{dH}
\end{equation}
by $\displaystyle{(W(\hat r_1+\varepsilon) W(\hat
r_1-\varepsilon))^{-1}\frac{dT}{dH}}$
multiplied by
\begin{eqnarray}\label{massivedothatr1}
&& R'\left[- W(\hat r_1-\varepsilon) (R'-1)+ W(\hat r_1+\varepsilon) (R'-1 
+ ( R - \hat r_1)\frac{\partial v_1}{\partial R}+(
R'- 1)\frac{\partial v_1}{\partial R'})\right]- \nonumber\\
&& (R- \hat r_1)\left[-W(\hat r_1-\varepsilon )R''+W(\hat r_1+\varepsilon)
(R''+ \frac{\partial v_1}{\partial R}R'+
\frac{\partial v_1}{\partial R'}R'')\right]
\end{eqnarray}
where $R,R',R''$ stay for $R(\hat r_1),R'(\hat r_1+\varepsilon),R''(\hat
r_1+\varepsilon)$.
From
\begin{equation}
W(\hat r_1-\varepsilon)=W(\hat r_1+\varepsilon) +\frac{\hat p_1}{R(\hat
r_1)}
\end{equation}
we find
\begin{equation}\label{dv1sdR1}
W(\hat r_1+\varepsilon) \frac{\partial v_1}{\partial R'(\hat r_1+\varepsilon)}
=\frac{\hat p_1}{R(\hat r_1)}
\end{equation}
which substituted into eq.(\ref{massivedothatr1}) makes it vanish. This 
is an expected
result as the motion of the exterior shell must not depend on the 
dynamics which
develops at smaller radiuses, but only on the two masses $H$ and $M_0$.

With regard to the terms proportional to $\dot{\hat r}_2$ in addition to
\begin{equation}
\dot{\hat r}_2 \frac{\partial p^0_{c2}}{\partial H}
\end{equation}
we have the term given by 
\begin{equation}
\frac{1}{W(\hat r_1+\varepsilon) W(\hat r_1-\varepsilon)} \frac{dT}{dH}
\end{equation}
multiplied by  
\begin{eqnarray}
&&(-(R'-1)+(R-\hat r_1)\frac{\partial T}{\partial \hat r_2})\bigg[
-W(\hat r_1-\varepsilon)(R'-1)+ \\
&& \left.W(\hat r_1+\varepsilon)(R'-1+\frac{\partial v_1}{\partial R}
(R-\hat r_1) + \frac{\partial v_1}{\partial R'}(R'-1))\right]- \nonumber\\
&& (R-\hat r_1)\bigg[-W(\hat r_1-\varepsilon)(\frac{dT}{d\hat r_2}(R'-1)-R'')+
W(\hat r_1+\varepsilon)(\frac{dT}{d\hat r_2}(R'-1)-R''+ \nonumber\\
&& \frac{\partial v_1}{\partial R}(-R'+1+\frac{dT}{d\hat r_2}( R-\hat r_1))+
\frac{\partial v_1}{\partial R'}(-R''+\frac{dT}{d\hat r_2}(R'-1)))\bigg]
\nonumber
\end{eqnarray}
where again $R,R',R''$ stay for $R(\hat r_1),R'(\hat
r_1+\varepsilon),R''(\hat r_1+\varepsilon)$.
As a consequence of eq.(\ref{dv1sdR1}) the above expression vanishes. 
Adding the contribution of the boundary term we have
\begin{equation}
\dot{\hat r}_2 \frac{\partial p^0_{c2}}{\partial H}-N(r_m) =0
\end{equation}
which is the single shell equation of motion (\ref{outereqmotion}).

The equation of motion of the interior shell is obtained from the variation
$\delta H=0$, $\delta M_0\neq 0$. In this case we have no contribution form the
boundary terms. For the coefficient of $\dot{\hat r}_1$ we
obtain 
\begin{eqnarray}\label{dotr1massive}
&& -\frac{1}{W(\hat r_1+\varepsilon)}+ R' R 
\left[-\frac{1}{W(\hat r_1+\varepsilon)} \frac{dT}{d M_0}(R'-1)+\right.\\
&&\left.\frac{1}{W(\hat r_1-\varepsilon)}\left(\frac{dT}{d M_0}(R'-1+
\frac{\partial v_1}{\partial R}(R-\hat r_1)+ \frac{\partial v_1}{\partial
R'}(R'-1))+ 
\frac{\partial v_1}{\partial M_0}\right)\right]- \nonumber\\
&& (R-\hat r_1)\frac{dT}{d M_0} R \left[-\frac{R''}{W(\hat
r_1+\varepsilon)}+\frac{1}{W(\hat r_1-\varepsilon)}(R''
+\frac{\partial v_1}{\partial R} R'+ 
\frac{\partial v_1}{\partial R'}R'')\right].\nonumber
\end{eqnarray}
having used the relation (see eq.(\ref{Fproperties}))
\begin{equation}
R'\frac{\partial^2 F}{\partial {\cal M}\partial R'}-
\frac{\partial F}{\partial {\cal M}}= -\frac{1}{W}~.
\end{equation}
Substituting in eq.(\ref{dotr1massive}) the relation
\begin{equation}
\frac{\partial v_1}{\partial M_0}=\frac{\hat p_1 (1+\frac{\hat p_1}
{R ~W(\hat r_1+
\varepsilon)})}{R(R'(\hat r_1+\varepsilon) \hat p_1 - V_1 W(\hat
r_1+\varepsilon))} 
\end{equation}
such a coefficient of $\dot{\hat r}_1$ becomes simply
\begin{equation}
\frac{1}{\frac{\hat p_1}{V_1}R'(\hat r_1+\varepsilon)-W(\hat r_1+\varepsilon)}
\end{equation}
which is the non zero mass generalization of the
coefficient of $\dot{\hat r}_1$ appearing in  eq.(\ref{firstdothatr1}).
Then using eqs.(\ref{NrNequations},\ref{Nrsolution}) for $N(\hat r_1)$ 
and $N^r(\hat r_1)$
we have 
\begin{eqnarray}
& & 0=\left[\frac{\hat p_1}{V_1}R'(\hat r+\varepsilon)-W(\hat
r+\varepsilon)\right]^{-1} 
\left(\dot{\hat r}_1 -\frac{\hat p_1}{V_1}N(\hat r_1)+N^r(\hat r_1)
\right) \nonumber\\
& & +\dot{\hat r}_2\left[\frac{\partial p^0_{c2}}{\partial
M_0}-\frac{\partial^2 
F}{\partial M_0\partial R'}\frac{V_2}{\hat r_2}+\frac{\partial F}
{\partial M_0}\right] 
+\alpha \\
& & - \left[\frac{\hat p_1}{V_1}R'(\hat r+\varepsilon)-W(\hat r
+\varepsilon)\right]^{-1}\frac{\hat p_1}{V_1}\frac{\dot R(\hat
r_1+\varepsilon)} {W(\hat r+\varepsilon)} \nonumber\\
& & +\dot{\hat r}_2\frac{\partial \tilde p_{c2}}{\partial M_0}-\dot R(\hat
r_1+\varepsilon)\frac{\partial^2 F}{\partial M_0\partial R'}(\hat
r_1+\varepsilon) -
\dot{\hat r}_2 \frac{d}{d\hat r_2}\left[(R(\hat r_1)-\hat r_1)\frac{dT}{dM_0}
{\cal D}\right] 
\nonumber
\end{eqnarray}
where $\alpha = \sqrt{2H\hat r}/(\sqrt{2H\hat r}+\hat p)$.
The first line is simply the equation of motion for $\hat r_1$, the second line
vanishes being simply the equation of motion for $\hat r_2$ 
due to the variation of $M_0$ (see eq.(\ref{generalMeqmotion})) while 
the sum of the third and fourth line vanishes in virtue of relation
(\ref{dv1sdR1}). 

\eject

\vfill

\end{document}